\documentstyle[11pt]{article}
\begin{document}
\title{\large\bf Cosmological production of charged black hole pairs}
\author{R.B. Mann$^a$ and Simon F. Ross$^b$ \\
Department of Applied Mathematics and Theoretical Physics \\
University of Cambridge, Silver St., Cambridge CB3 9EW \\
$^a${\it R.B.Mann@damtp.cam.ac.uk} \\ $^b${\it S.F.Ross@damtp.cam.ac.uk}}
\date{\today \\ DAMTP/R-95/9}
\maketitle
\begin{abstract}
We investigate the pair creation of charged black holes in a background
with a positive cosmological constant. We consider $C$ metrics with a
cosmological constant, and show that the conical singularities in the
metric only disappear when it reduces to the Reissner-Nordstr\"om de
Sitter metric. We construct an instanton describing the pair
production of extreme black holes and an instanton describing the
pair production of non-extreme black holes from the
Reissner-Nordstr\"om de Sitter metric, and calculate their
actions. There are a number of striking similarities between these
instantons and the Ernst instantons, which describe pair production in
a background electromagnetic field. We also observe that the type I
instanton in the ordinary $C$ metric with zero cosmological constant
is actually the Reissner-Nordstr\"om solution.
\end{abstract}

\pagebreak

\section{Introduction}
\label{intro}

There has been considerable interest recently in studying black hole
pair creation by instanton methods, and a number of interesting
results have been obtained \cite{dgkt,dggh,entar}. The study of black
hole pair creation has so far been mostly restricted to the pair
creation of oppositely-charged black holes by an electromagnetic
field, where pair creation is possible because the negative potential
energy of the created pair of black holes in the background
electromagnetic field balances their rest mass energy. Black holes
can, however, also be pair created by a cosmological background, as a
positive cosmological constant supplies the necessary negative
potential energy. We will examine the pair creation of electrically or
magnetically charged black holes in a cosmological background. There
are two instantons, which describe pair production of non-extreme and
extreme black holes respectively. There is another instanton, the
charged generalisation of the instanton found in \cite{desit}, but
careful analysis suggests this latter instanton does not formally represent
black hole pair creation.

We find that some of the most interesting results of the
electromagnetic case can be reproduced in the cosmological case; in
particular, the pair creation rate is still determined by the entropy
of the solutions. Indeed, there is an interesting similarity between
the instantons we find and the instantons found in
\cite{dgkt,dggh} for the pair creation of black holes in an
electromagnetic field. We feel that these similarities encourage the
idea that the results of these instanton calculations represent real
quantum gravity effects, and will not be qualitatively modified by the
inclusion of quantum corrections. In other words, since the main features are
similar in these two disparate models, we think that they represent
features that should also be present in the full quantum theory.

We will begin by considering the charged $C$ metric, which can be
interpreted as representing a pair of oppositely-charged black holes
accelerating away from each other in a flat background spacetime. As
there is no force to accelerate the black holes, this metric is not in
general regular. The approach taken in \cite{ernst,dgkt} was to add a
background electromagnetic field to this metric by a Harrison
transformation, giving the Ernst metric, which could be made regular
by a suitable choice of field strength. The background field provides
the necessary force to accelerate the black holes. We will consider
adding a cosmological constant to the $C$ metric instead, and see if
we can obtain a non-singular solution by using the cosmological
acceleration to accelerate the black holes. We find that we can, but
that the $C$ metric reduces to the Reissner-Nordstr\"om de Sitter
solution whenever the acceleration of the black holes is matched to
the cosmological acceleration in this way.

There are in fact two ways to obtain a non-singular solution in the
charged $C$ metric itself. One is to set the acceleration to zero, in
which case the $C$ metric reduces to the Reissner-Nordstr\"om metric.
The other way to get rid of the conical singularities is to allow the
black hole and acceleration horizons to coincide, which has been
referred to as the Type I instanton \cite{dgkt}. We have found that
the $C$ metric in this special case has the same functional form as
the Reissner-Nordstr\"om metric, but it is now what is usually
regarded as the azimuthal coordinate which is playing the role of
time. That is, one can obtain this metric by analytically continuing
$\phi \to i \Phi$ in the Euclidean Reissner-Nordstr\"om metric.

As the only non-singular versions of the charged $C$ metric with a
cosmological constant reduce to the Reissner-Nordstr\"om de Sitter
metric, we would like to know what instantons can be made from this
metric. We find that there are four types of instantons, referred to
as the lukewarm, cold, ultracold \cite{romans}, and charged Nariai
solutions. The Euclidean section of the lukewarm
solution has topology $S^2 \times S^2$, and it can be
thought of as representing pair creation of non-extreme black
holes. The Euclidean section of the cold solution has topology $S^2
\times R^2$, and it can be thought of as representing pair
creation of extreme black holes (by an extreme black hole, we mean one
in which the inner and outer horizons coincide).  The charged Nariai
solution has topology $S^2 \times S^2$; its interpretation is not
clear. The ultracold solution is just a special case of the cold
solution where all three horizons coincide, but the metric is quite
different in this special case. There is an analogy between this set
of instantons and the instantons constructed from the Ernst solution
in \cite{dggh}: there one had an extreme black hole instanton, a
non-extreme black hole instanton, and the Type I instanton; here we
have the cold, lukewarm and charged Nariai instantons.

In the instanton approach, the wavefunction is approximated by $\Psi
\approx e^{-I}$, where $I$ is the action of the instanton, so the
partition function, which gives the pair creation rate, is
approximated by $Z \approx e^{-2I}$. The action for
Einstein-Maxwell with a cosmological constant is
\begin{equation} \label{emac}
I = -\frac{1}{16\pi} \int
d^4x\sqrt{g}\left(R-2\Lambda-F_{\mu\nu}F^{\mu\nu}\right) +
\frac{1}{8\pi} \int_\Sigma d^3x \sqrt{h} K,
\end{equation}
where $R$ is the Ricci scalar of the metric $g_{\mu\nu}$, $\Lambda$ is
the cosmological constant (we assume $\Lambda >0$), and $F_{\mu\nu}$
is the electromagnetic field tensor. The boundary of the manifold is
$\Sigma$, which has metric $h_{ij}$, and extrinsic curvature
$K_{ij}$. Since there is no asymptotic region for these solutions,
this action is the microcanonical action \cite{hawross}, and thus the
entropy is \cite{brown}
\begin{equation}
S = \log Z = -2I.
\end{equation}

In section \ref{action}, we calculate the action for the instantons
described in section \ref{instsec}. We find that the pair creation rate
given by these instantons is smaller than the rate which de Sitter
space gives to propagate from nothing to an $S^3$. Thus the
contribution from de Sitter space will dominate that of these
instantons in the path integral, in agreement with experience. We also find
that the entropy of these solutions is always equal to one quarter of
the area of the horizons which appear on the Euclidean section, as
expected. These instantons are thus very similar to the Ernst
instantons. We summarise our results in section \ref{concl}.

\section{Cosmological $C$ metrics}
\label{cosmoc}

The well-known $C$ metric solution of the Einstein-Maxwell equations,
which describes a pair of charged black holes undergoing uniform
acceleration, was found in \cite{Cmetric}. There is a less well-known
generalisation of this solution to include a cosmological constant
\cite{gen}. In the usual $C$ metric, it is not possible to eliminate
the possible conical singularities at both poles in general. The $C$
metric is interpreted as describing two oppositely-charged black holes
undergoing uniform acceleration, and these singularities are
interpreted as representing ``rods'' or ``strings'' which provide the
force necessary to accelerate the black holes.

If we consider the $C$ metric with a positive cosmological constant,
one might think that the cosmological constant can provide the
necessary force, and so we should be able to eliminate the conical
singularities, and thus obtain a completely regular metric. What we
find is that the elimination of the conical singularities, which we
regard as fixing the acceleration parameter $A$, causes the metric to
reduce to the Reissner-Nordstr\"om de Sitter metric.

The cosmological $C$ metric can be written as
\begin{equation} \label{ccmet}
ds^2 = \frac{1}{A^2 (x-y)^2} \left[ H(y) dt^2 - H^{-1}(y) dy^2 +
G^{-1}(x) dx^2 + G(x) d\varphi^2 \right],
\end{equation}
where
\begin{equation}
G(x) = 1 -x^2(1+r_+ A x)(1+r_- Ax)- \frac{\Lambda}{3 A^2},
\end{equation}
and
\begin{equation}
H(y) = 1-y^2 (1+r_+ Ay) (1+r_- Ay).
\end{equation}
The cosmological constant is $\Lambda$, which we will assume is
positive.  The gauge field in the magnetic case is
\begin{equation} \label{gauge1}
F = -q dx \wedge d\varphi,
\end{equation}
and the gauge field in the electric case is
\begin{equation} \label{gauge2}
F = -q dt \wedge dy,
\end{equation}
where $q=\sqrt{r_+ r_-}$. We will denote by $x_1,x_2,x_3, x_4$ the roots of
$G(x)$ in ascending order, and by $y_1,y_2, y_3, y_4$ the roots of
$H(y)$ in ascending order. We will restrict the parameters $\Lambda,
r_+, r_-$ and $A$ so that all these roots are real.  As with the usual
$C$ metric, $y=y_1$ is interpreted as the inner black hole horizon,
$y=y_2$ is interpreted as the outer black hole horizon, and $y=y_3$ is
interpreted as the acceleration or cosmological horizon.

We will first restrict $x$ to $x_3 \leq x \leq x_4$ in order to obtain
a metric of the appropriate signature.  We will restrict $y$ to
$-\infty < y \leq x$, as when $y=x$ the conformal factor in the metric
diverges, and so this corresponds to infinity. In order to have a
regular solution, we must avoid having conical singularities at
$x=x_3$ or $x=x_4$, so we must demand that \cite{dgkt}
\begin{equation} \label{nostrut}
G'(x_3) =- G'(x_4),
\end{equation}
and identify $\varphi$ periodically with period $\Delta \varphi =
4\pi/ |G'(x_3)|$. This assumes that $x_3$ is at a finite distance
from $x_4$, but if one took $x_3 = x_2$, then $x=x_3$ would lie at infinite
proper distance from any other point, so that there could be no
conical singularity there. That is, the $(x,\varphi)$ sections would no
longer be compact. This is analogous to what happens with the
Euclidean section of an extreme black hole \cite{entar}.  However, for
positive cosmological constant, $H(y)$ will have just two real roots
if $x_2=x_3$, so this contradicts our restriction of the
parameters. Therefore we must satisfy (\ref{nostrut}), which
corresponds to
\begin{equation}
(x_3-x_4) (x_3-x_2) (x_3-x_1) = (x_3-x_4) (x_4-x_2)(x_4-x_1),
\end{equation}
and can only be satisfied by taking $x_3=x_4$.

This seems to imply that the $(x,\varphi)$ section shrinks to a point,
but this is just due to a poor choice of coordinate system. In the
limit that $x_3 =x_4$, the proper distance between $x_3$ and $x_4$
remains finite, as can be seen by the following coordinate
transformation.\footnote{This coordinate transformation is inspired by
that in \cite{desit}.} The limit $x_3 \to x_4$ corresponds to the
limit $A \to \sqrt{\Lambda /3}$ from above. Let us write $1 - \Lambda
/(3A^2) = \epsilon^2$, so that the appropriate limit is $\epsilon \to
0$. With this parametrisation,
\begin{equation}
G(x) = \epsilon^2 - x^2(1+r_+Ax) (1+r_-Ax),
\end{equation}
so $x_3 \approx -\epsilon$ and $x_4 \approx \epsilon$. Then set
\begin{equation}
x = \epsilon \cos \theta, \varphi = \frac{\phi}{\epsilon}.
\end{equation}
In the limit $\epsilon \to 0$, the metric becomes
\begin{equation}
ds^2 = \frac{3}{\Lambda y^2} \left[ H(y) dt^2 - H^{-1}(y) dy^2 + d\theta^2
+ \sin^2 \theta d\phi^2 \right],
\end{equation}
and if we make the further coordinate transformation
\begin{equation}
r = -\frac{\sqrt{3/\Lambda}}{y}, T =  \sqrt{3/\Lambda} t,
\end{equation}
 it becomes
\begin{equation}
ds^2 = -V(r) dT^2 + \frac{dr^2}{V(r)} + r^2(d\theta^2
+\sin^2 \theta d\phi^2),
\end{equation}
where
\begin{equation}
V(r) = \left(1-\frac{r_+}{r}\right) \left( 1
-\frac{r_-}{r}\right) - \frac{\Lambda r^2}{3}.
\end{equation}
Note also that as $\varphi$ has period $4 \pi/|G'(x_3)|$, $\phi$ has
period $2\pi$. The range of $r$ is from $0$ to $\infty$, and
$\theta$ runs from $0$ to $\pi$. The gauge field is now
\begin{equation}
 F =  q \sin \theta d\theta \wedge d\phi
\end{equation}
in the magnetic case, and
\begin{equation}
F = -\frac{q}{r^2} dT \wedge dr
\end{equation}
in the electric case.  Thus, this can be identified as a
Reissner-Nordstr\"om de Sitter solution with charge $q =
\sqrt{r_+r_-}$ and `mass' $M = (r_+ + r_-)/2$. Note that if we let
$\Lambda=0$, the limit $x_3=x_4$ is just the limit $A\to 0$, and what
we have just done specialises to the familiar statement that the $C$
metric reduces to the Reissner-Nordstr\"om metric in this limit
\cite{dgkt}.

We have assumed earlier that $-\infty < y < x$ to avoid divergence in
the conformal factor, but we could equally well have taken $x < y <
\infty$. Then $y=y_2$ is interpreted as the cosmological horizon,
$y=y_3$ is the outer black hole horizon, and $y=y_4$ is the inner
black hole horizon. We now have to take $x_1 \leq x \leq x_2$ to get a
metric of the right signature.

To make this metric regular, we must take $\varphi$ to be periodic
with period $\Delta \varphi = 4 \pi/ |G'(x_1)|$, and require
\begin{equation}
G'(x_1) = -G'(x_2).
\end{equation}
This implies
\begin{equation}
(x_1 -x_2)(x_1-x_3)(x_1-x_4) = (x_1-x_2)(x_2-x_3)(x_2-x_4),
\end{equation}
and thus $x_1= x_2$. We have found that this occurs when $A=A_c$, where
\begin{equation} \label{seta}
A_c^2 = \frac{\Lambda}{3} - \frac{[3(r_+ + r_-)+\gamma]^2
[(r_++r_--\gamma)^2-16 (r_+-r_-)^2]}{4096 r_+^3 r_-^3},
\end{equation}
and
\begin{equation}
\gamma = \sqrt{9(r_+^2+r_-^2)-14r_+r_-}.
\end{equation}
if we let $ A_c^2/ A^2 = 1 - \epsilon^2$, and
$X=x-x_1$, then
\begin{equation}
G(x) = \epsilon^2 - \frac{X^2}{16r_+r_-}(a + b AX + c A^2 X^2),
\end{equation}
where
\begin{equation}
a = (3(r_+ + r_-) + \gamma) \gamma,
\end{equation}
\begin{equation}
b = - 8 r_+r_-(r_-+r_++\gamma),
\end{equation}
and
\begin{equation}
c = 16r_+^2r_-^2.
\end{equation}
Therefore, if we make coordinate transformations
\begin{equation}
X = \sqrt{\frac{16r_+r_-}{a}} \epsilon \cos \theta,
\end{equation}
and
\begin{equation}
\varphi = \sqrt{\frac{16r_+r_-}{a}} \phi/\epsilon,
\end{equation}
the metric becomes, in  the limit $\epsilon \to 0$,
\begin{equation}
ds^2 = \frac{1}{A_c^2 (y-x_1)^2}\left[ H(y) dt^2 - H^{-1}(y) dy^2 +
\frac{16r_+r_-}{a} (d\theta^2 + \sin^2 \theta d\phi^2)\right].
\end{equation}
If we make the further coordinate transformations
\begin{equation}
r = \sqrt{\frac{16r_+r_-}{a}}\frac{1}{A_c(y-x_1)}, T =
\sqrt{\frac{a}{16 r_+r_-}}  \frac{t}{A_c},
\end{equation}
the metric will become
\begin{equation}
ds^2 = -V(r)dT^2 + \frac{dr^2}{V(r)} + r^2 (d\theta^2 +
\sin^2 \theta d\phi^2),
\end{equation}
where
\begin{equation}
V(r) = 1 -\frac{2M}{r} + \frac{Q^2}{r^2} -\frac{\Lambda}{3} r^2,
\end{equation}
and
\begin{equation}
M =-\sqrt{\frac{16r_+r_-}{a}}\frac{b}{2a} ,
\qquad  Q^2= \frac{16 r_+r_- c}{a^2}.
\end{equation}
The gauge field becomes
\begin{equation}
F = \frac{16 r_+r_-}{a}q \sin \theta d\theta \wedge d\phi = Q \sin
\theta d\theta \wedge d\phi
\end{equation}
in the magnetic case, and
\begin{equation}
F = \frac{Q}{r^2} dT \wedge dr
\end{equation}
in the electric case.
Therefore, this can be identified as a Reissner-Nordstr\"om
de Sitter metric as well. Note that although the equations are more
complicated, $M$ and $Q$ are still just functions of $r_+$ and
$r_-$, and $M$ is positive if $r_+$ and $r_-$ are positive.

If the cosmological constant is set to zero, we see that the $C$
metric is again only non-singular when it reduces to the
Reissner-Nordstr\"om metric, that is, when the acceleration of the
black holes vanishes. However, in this case, this happens for non-zero
$A$, as we can easily see from (\ref{seta}). This is just an
indication that we shouldn't think of $A$ as simply parametrising the
acceleration in this case.

\section{The Type I instanton}
\label{type1}

When the cosmological constant is set to zero, we have seen that the
ordinary $C$ metric has no conical singularities when the acceleration
vanishes, as we should have expected.  However, when $\Lambda=0$, we
can eliminate the conical singularities in another way, as was first
observed by Dowker {\it et al} \cite{dgkt}. It is now possible to set
$x_3=x_2$, as $x_3=y_3=\xi_3$ and $x_2=y_2=\xi_2$.  Again, the
apparent degeneracy of the two roots is merely an artifact of our
coordinate system. As explained in \cite{dgkt}, we may make a
transformation so that the metric remains regular when $\xi_3 =
\xi_2$.

The $C$ metric is given by (\ref{ccmet}) with $\Lambda=0$. However,
for consistency with \cite{dgkt}, we will adopt a slightly different
coordinate system in this section, and write the $C$ metric as
\begin{equation}
ds^2 = \frac{1}{A^2(x-y)^2} \left[ G(y) dt^2 - G^{-1}(y) dy^2 +
G^{-1}(x) dx^2 + G(x) d\varphi^2 \right]
\end{equation}
with
\begin{equation}
G(\xi) = [1-\xi^2(1+r_+A\xi)](1+r_- A \xi).
\end{equation}
The transformation between the two forms of the $C$ metric is
discussed in \cite{Cmetric}.  The gauge field is still (\ref{gauge1})
or (\ref{gauge2}). We will refer to the roots of $G(\xi)$ as $\xi_1,
\xi_2, \xi_3, \xi_4$ in ascending order. As before, $x$ is restricted
to $\xi_3 \leq x \leq \xi_4$ to obtain the appropriate signature.  In
fact, if $\xi_2 = \xi_3$, the appropriate range is actually $\xi_3 < x
\leq \xi_4$, and the $(x, \varphi)$ section becomes topologically
$R^2$. We can then obtain a regular solution by identifying $\varphi$
with period $4\pi/ |G'(\xi_4)|$.

The two roots $\xi_3$ and $\xi_4$ will coincide when $A = A_c =
2/(3\sqrt{3}r_+)$. Following \cite{dgkt}, we let $r_+ A = 2/(3{\sqrt
3}) - \epsilon^2/{\sqrt 3}$, so that the limit of coincident roots is
$\epsilon \rightarrow 0$. If we make the coordinate transformation
\begin{equation}
y = \sqrt{3}(-1+\epsilon \cos \chi), \quad \psi = \sqrt{3} \epsilon t
\end{equation}
the metric is, in the limit $\epsilon \to 0$,
\begin{equation}
ds^2 = \frac{1}{A_c^2(x+\sqrt{3})^2} \left[ -\alpha\sin^2 \chi d\psi^2 +
\alpha^{-1} d\chi^2 + G^{-1}(x) dx^2 + G(x) d\varphi^2 \right],
\end{equation}
where now
\begin{equation}
G(x) =-\frac{2}{3\sqrt{3}}(x+\sqrt{3})^2 (x-\sqrt{3}/2) \left(
1 + \frac{2 r_- x}{3 \sqrt{3} r_+} \right)
\end{equation}
and
\begin{equation}
\alpha = 1 - \frac{2r_-}{3r_+}.
\end{equation}
If we analytically continue $\psi \to i \Psi$, and identify $\Psi$
with period $2\pi \alpha^{-1}$, we obtain an instanton with topology
$S^2 \times R^2$, referred to as the type I instanton. It was not
initially clear how to interpret this instanton. However, our experience
with the cosmological $C$ metric above suggests that it is related to
the Reissner-Nordstr\"om instanton. We are encouraged in this guess
by the fact that the $(\Psi, \chi)$ two-sphere sections are round.

The range of $x$ in this solution is $-\sqrt{3} < x \leq
\sqrt{3}/2$. Let $\tilde{x} = x+\sqrt{3}$, and $\phi = \alpha
\Psi$, so that $\phi$ has period $2\pi$, and the Euclideanised metric
becomes
\begin{equation}
ds^2 = \frac{\alpha^{-1}}{A_c^2 \tilde{x}^2} \left[ d\chi^2 + \sin^2\chi
d\phi^2 + \alpha ( G^{-1}(\tilde{x}) d\tilde{x}^2 + G(\tilde{x})
d\varphi^2)\right].
\end{equation}
If we make a further coordinate transformation,
\begin{equation}
r = \frac{\alpha^{-1/2}}{A_c \tilde{x}}, \tau = \frac{\alpha^{1/2}
\varphi}{A_c},
\end{equation}
the metric becomes
\begin{equation} \label{ern}
ds^2 = r^2(d\chi^2+ \sin^2 \chi d\phi^2) + V(r) d\tau^2 + \frac{dr^2}{V(r)},
\end{equation}
where
\begin{equation} \label{Vr}
V(r) = \left( 1 - \frac{\tilde{r}_+}{r} \right) \left( 1 -
\frac{\tilde{r}_-}{r} \right).
\end{equation}
The new parameters are $\tilde{r}_+ = r_+ \alpha^{-1/2}$ and
$\tilde{r}_- = - r_- \alpha^{-3/2}$. We also note that $r$ runs from
$\tilde{r}_+$ to $\infty$ on the Euclidean section. If the gauge
field is (\ref{gauge1}), it becomes
\begin{equation}
F = \frac{q \alpha^{-1}}{r^2} dr \wedge d\tau =
-\frac{iQ}{r^2} d\tau \wedge dr,
\end{equation}
and if it is (\ref{gauge2}), it becomes
\begin{equation}
F = -iq \alpha^{-1} \sin \chi d\chi \wedge d\Psi = Q \sin \chi d\chi
\wedge d\phi,
\end{equation}
where $Q^2 = \tilde{r}_+ \tilde{r}_-$. Therefore, this instanton is
seen to be the Euclidean Reissner-Nordstr\"om solution with charge $Q$
and mass $M = \frac{1}{2} (\tilde{r}_+ + \tilde{r}_-)$, with the
magnetic instanton being identified with the electric
Reissner-Nordstr\"om solution and vice-versa. Thus we see again that
eliminating the conical singularities is only possible when the $C$ metric
reduces to Reissner-Nordstr\"om.

However, the coordinate we analytically continued to obtain a
Euclidean section has been identified with the azimuthal coordinate in
the Euclidean Reissner-Nordstr\"om solution. This means that the
problem of understanding the physical significance, if any, of the $C$
metric in this special case is equivalent to giving a physical
interpretation of the metric (\ref{ern}) with $\phi = i\Phi$. Because
$\tau$ is not analytically continued, this metric has one compact
direction. This leads us to suspect that the interpretation is similar
to that given for the five-dimensional black holes in \cite{witten};
that is,  that this instanton should be interpreted as
representing the decay of a Kaluza-Klein vacuum, ${\rm Mink}^{(2,1)}
\times S^1$.

The analytically continued metric is
\begin{equation} \label{newm}
ds^2 =  - r^2 \sin^2 \chi d\Phi^2 + r^2 d\chi^2 + V(r) d\tau^2 +
\frac{dr^2}{V(r)},
\end{equation}
where $V(r)$ is given in (\ref{Vr}). So long as the period $\Delta
\tau$ around the compact direction is chosen appropriately, the
coordinate singularity at $r=\tilde{r}_+$ is harmless. Let us ignore for the
moment the factors of $V(r)$, and consider just the three-dimensional
space $(r,\chi,\Phi)$, that is, consider
\begin{equation} \label{3metric}
ds^2 = -r^2 \sin^2 \chi d \Phi^2 + dr^2 + r^2 d\chi^2.
\end{equation}
This metric describes a portion of three-dimensional Minkowski
space. In fact, if we make the coordinate transformations
\begin{equation} \label{coord1}
z = r \sin \chi \cosh \Phi,\quad  t = r\sin\chi \sinh \Phi,\quad y = r
\cos \chi
\end{equation}
the metric (\ref{3metric}) becomes
\begin{equation}
ds^2 = -dt^2 + dz^2 + dy^2,
\end{equation}
the usual metric on Minkowski space. Because $\chi$ is restricted to
$0 \leq \chi \leq \pi$, the original coordinates only
cover the part $z \geq 0$ of the Minkowski space;  however there is no
obstruction to
extending the solution to the whole of Minkowski space. We could also set
\begin{equation} \label{coord3}
z = R \sin \psi, y = R \cos \psi,
\end{equation}
and rewrite this metric as
\begin{equation} \label{mink2}
ds^2 = -dt^2 + dR^2 + R^2 d\psi^2.
\end{equation}

Now let us consider the effect of the factors of $V(r)$ in
(\ref{newm}). The metric (\ref{newm}) approaches  the product manifold
${\rm Mink}^{(2,1)} \times S^1$ asymptotically, but the coordinate $r$
is now restricted to $\tilde{r}_+ \leq r \leq \infty$. We can still
make the same coordinate transformations,
(\ref{coord1},\ref{coord3}). The resulting metric
approaches (\ref{mink2})$\times S^1$ asymptotically, but the
restriction $r \geq \tilde{r}_+$ implies $R^2 - t^2 \geq
\tilde{r}_+^2$. Note that it is not possible to continue the metric
beyond $r=\tilde{r}_+$, as the radius of the circle direction vanishes
there.

The physical interpretation is exactly the same as in \cite{witten}:
if we assume the radius of the circle direction is relatively small,
putative observers living in this space who didn't go too close to $r=
\tilde{r}_+$ would think  (\ref{newm}) described three-dimensional
Minkowski space with the interior of the hyperboloid $R^2 -t^2 =
\tilde{r}_+^2$ omitted.  The type I instanton should be interpreted as
representing tunnelling from the vacuum (\ref{mink2}) cross a circle to
(\ref{newm}); that is, it describes the decay of a three-dimensional
Kaluza-Klein vacuum.

\section{Reissner-Nordstr\"om de Sitter instantons}
\label{instsec}

Since the only special cases of the cosmological $C$ metrics for which
the metric is regular reduce to the Reissner-Nordstr\"om de Sitter
metrics,  consideration of the pair creation of charged black holes
in a background with a positive cosmological constant reduces to a
consideration of the non-singular instantons that can be constructed
from the Reissner-Nordstr\"om de Sitter metric. This question has of
course been studied before, notably in \cite{romans,moss}, and in the
uncharged case in \cite{desit}, but we hope to present a unified
picture which makes the relations between the instantons clear.

The Reissner-Nordstr\"om de Sitter metric is
\begin{equation} \label{RNS}
ds^2 = -V(r) dt^2 + \frac{dr^2}{V(r)} + r^2 (d\theta^2 + \sin^2 \theta
d \phi^2 ),
\end{equation}
where
\begin{equation}
V(r) = 1 - \frac{2M}{r} + \frac{Q^2}{r^2} - \frac{1}{3} \Lambda r^2.
\end{equation}
The gauge field is
\begin{equation}
F = -\frac{Q}{r^2} dt \wedge dr \label{el1}
\end{equation}
for an electrically-charged solution, and
\begin{equation}
F = Q \sin \theta d\theta \wedge d\phi \label{mag1}
\end{equation}
for a magnetically-charged solution. For the sake of simplicity, we
will not consider dyonic solutions. This solution has three
independent parameters, the `mass' $M$, charge $Q$ and cosmological
constant $\Lambda$, which we will assume are all positive. There are
then four roots of $V(r)$, which we designate by $r_1, r_2, r_3, r_4$
in ascending order. The first root is negative, and therefore has no
physical significance. The remaining roots are interpreted as various
horizons; $r=r_2$ is the inner (Cauchy) black hole horizon, $r=r_3$ is
the outer (Killing) black hole horizon, and $r=r_4$ is the
cosmological (acceleration) horizon. In the Lorentzian section, $0 \leq r
< \infty$.

We want to construct instantons from this metric by analytically
continuing $t \to i\tau$. To obtain a positive-definite metric, we
must restrict $r$ to $r_3 \leq r \leq r_4$. There is then potentially
a conical singularity at $r=r_3$ and at $r=r_4$. However, if
$r_3=r_2$, the range of $r$ in the Euclidean section will be $r_3 < r
\leq r_4$, as the double root in $V(r)$ implies that the proper
distance from any other point to $r=r_3$ along spacelike directions is
infinite. In this case, we may obtain a regular instanton by
identifying $\tau$ periodically with period $2\pi / \kappa_4$, where
$\kappa_4$ is the surface gravity of the horizon $r=r_4$. This
instanton will be referred to as the cold instanton, following
\cite{romans}.

If we do not have $r_2 = r_3$, then we must have
\begin{equation} \label{permat}
\kappa_3 = \kappa_4,
\end{equation}
 and identify $\tau$ with the same period. There are two ways to
satisfy this condition; one is $r_3 = r_4$, which gives an instanton
analogous to that constructed out of the Nariai metric in
\cite{desit}, which we shall refer to as the charged Nariai
instanton. This is very similar to the way in which the Type I
instanton is obtained. The other is to set $Q = M$, which implies
(\ref{permat}) \cite{moss}; we shall refer to this as the lukewarm
instanton, following \cite{romans}. There is also a special case, when
$r_2 = r_3 = r_4$, which we refer to as the ultracold instanton, again
following \cite{romans}.

The form of all these instantons in terms of the metric (\ref{RNS})
has been given in detail in \cite{romans}; we will briefly summarise
that discussion here. If there is a double root $\rho$ of $V(r)$,
we can write
\begin{equation}
V_d (r) = \left( 1 - \frac{\rho}{r} \right)^2 \left( 1 -
\frac{1}{3}\Lambda (r^2 + 2 \rho r + 3 \rho^2) \right),
\end{equation}
and the mass and charge are thus given by
\begin{equation} \label{mrho}
M = \rho \left(1 - \frac{2}{3}\Lambda \rho^2 \right),
\end{equation}
\begin{equation} \label{qrho}
Q^2 = \rho^2 (1 -\Lambda \rho^2).
\end{equation}
For positive $\Lambda$, the double root $\rho$ must lie in $0 < \rho <
\Lambda^{-1/2}$. The other positive root of $V_d(r)$ is
\begin{equation}
b= \sqrt{3 \Lambda^{-1} - 2\rho^2} - \rho.
\end{equation}
For $0<\rho^2 < \Lambda^{-1}/2$, $b>\rho$, so $r_2=r_3=\rho\;$: this
solution gives the cold instanton. For $\Lambda^{-1}/2 < \rho^2 <
\Lambda^{-1}$, $b<\rho$, so $r_3=r_4=\rho\;$: this solution gives the
charged Nariai instanton. If $\rho^2 =\Lambda^{-1}/2$, then
$b=\rho=r_2=r_3=r_4\;$: this solution gives the ultracold instanton. The
function $V_d(r)$ can also be rewritten as
\begin{equation} \label{droot}
V_d(r) = \frac{-r^2}{(b^2+2\rho b+3\rho^2)} \left( 1 - \frac{\rho}{r}
\right)^2 \left(1 - \frac{b}{r} \right) \left( 1+ \frac{2\rho+b}{r} \right),
\end{equation}
and we could also write $M,Q$ and $\Lambda$ as functions of $\rho$ and
$b$ (see \cite{romans} for details).

If $V(r)$ does not have a double root, then it must have two roots
$r_3$ and $r_4$ such that
\begin{equation}
\kappa_3 = \frac{1}{2} |V'(r_3)| = \kappa_4
= \frac{1}{2} |V'(r_4)|.
\end{equation}
This fixes $V(r)$ to have the form
\begin{equation}
V_l (r) = \left( 1- \frac{r_3 r_4}{(r_3+r_4)r} \right)^2 -
\frac{r^2}{(r_3+r_4)^2},
\end{equation}
from which one sees immediately that $Q=M$. We will now comment
briefly on the nature of each of these instantons in turn.

For the lukewarm instanton, the topology of the Euclidean section is
$S^2 \times S^2$. The common temperature of the two horizons is \cite{moss}
\begin{equation}
T = \frac{1}{2\pi} \sqrt{\frac{\Lambda}{3} \left( 1 - 4M \sqrt{
\frac{\Lambda}{3}} \right)}.
\end{equation}
The Lorentzian section describes two black holes in de Sitter space,
so this instanton represents pair creation of non-extreme black holes in
thermal equilibrium with the cosmological acceleration radiation.
For the cold instantons, the horizon at $r=r_3$ is at infinite
distance, so the Euclidean section has topology $S^2 \times
R^2$. There is a boundary $B^{\infty}$ at the internal infinity
$r=r_3=\rho$. In the calculation of the action, we will take the
boundary to lie at $r = \rho + \epsilon$, and then take the limit as
$\epsilon \to 0$, that is, as the boundary approaches $B^\infty$. The
relation between the charge and mass is given parametrically by
(\ref{mrho},\ref{qrho}), and is displayed on Figure 1. The temperature
of the horizon at $r=r_4$ is \cite{romans}
\begin{equation}\label{tcold}
T = \frac{b}{2\pi (b^2+2\rho b + 3\rho^2)} \left (1 -\frac{\rho}{b}
\right)^2 \left( 1 + \frac{\rho}{b} \right).
\end{equation}
The Lorentzian section describes two extreme black holes in de Sitter space,
so this instanton represents pair creation of extreme black holes in
thermal equilibrium with the acceleration radiation.
For the charged Nariai instantons, where $r_3 = r_4$, it is necessary
to make a coordinate transformation and rewrite the metric as
\cite{hawross}
\begin{equation} \label{chN}
ds^2 = \frac{1}{A} (d\chi^2 + \sin^2 \chi d\psi^2) + \frac{1}{B}
(d\theta^2 + \sin^2 \theta d\phi^2),
\end{equation}
where $A$ and $B$ are constants, with $B>A$, $\chi$ and $\theta$ run from $0$
to
$\pi$, and $\psi$ and $\phi$ are periodic coordinates with period
$2\pi$. The gauge field becomes
\begin{equation}\label{elN}
F = Q \sin \theta d\theta \wedge d\phi
\end{equation}
in the magnetic case, and
\begin{equation}\label{magN}
F = -i Q \frac{B}{A} \sin \chi d\chi \wedge d\psi
\end{equation}
in the electric case. The cosmological constant is $\Lambda =
\frac{1}{2} (A+B)$, while $\rho^2 = 1/B$, and $M$ and $Q$ are given by
(\ref{mrho},\ref{qrho}).  This instanton has topology $S^2 \times
S^2$; indeed, it is just the direct product of two round two-spheres
with different radii.  The relation between the mass and charge is
still given parametrically by (\ref{mrho},\ref{qrho}), and is also
displayed on Figure 1. However, although this solution is just a
special case of Reissner-Nordstr\"om de Sitter, we find that the
singularity retreats to infinite proper distance when $r_3 =
r_4$, and so there is no longer a global event horizon, and the
Lorentzian section is just the direct product of two-dimensional de
Sitter space and a two-sphere of fixed radius, dS$^2 \times
S^2$. Thus the instanton doesn't represent pair creation of black
holes. However, as in \cite{desit}, higher-order quantum corrections
will break the degeneracy of the two roots, so the charged Nariai
solution will revert to an ordinary Reissner-Nordstr\"om de Sitter
spacetime once these effects are included.

\begin{figure}
\vspace{65mm}
\includegraphics{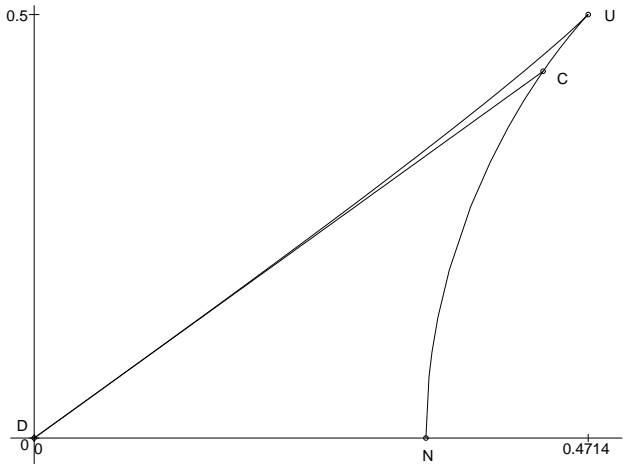}
\caption{The values of $Q$ and $M$ for which instantons can be
obtained in the cosmological case. The plot is of the dimensionless
quantities $Q\protect\sqrt{\Lambda}$
vs. $M\protect\sqrt{\Lambda}$. The curve DU represents the cold
solutions, DC represents the lukewarm solutions, and NU represents the
charged Nariai solutions. The point at D is de Sitter space, while U
is the ultracold case, and N is the Nariai solution. }
\end{figure}

The ultracold case deserves a more detailed description. In this case
$r_2=r_3=r_4$, which may be regarded as the limit $b=\rho$ in the
metric (\ref{RNS}) with $V(r)$ having the double root form
(\ref{droot}). In this case, the mass and charge are given by
(\ref{mrho},\ref{qrho}) with $\rho = 1/\sqrt{2\Lambda}$, that is,
\begin{equation}
M=  \frac{2}{3 \sqrt{2\Lambda}},\qquad  Q^2 = \frac{1}{4\Lambda}.
\end{equation}
Suppose $\rho = 1/\sqrt{2\Lambda} - \epsilon$, and $b =
1/\sqrt{2\Lambda} + \epsilon$, and consider the limit $\epsilon \to
0$. We can construct two different metrics in this limit. First,
define a new coordinate $R$ by
\begin{equation}
r = 1/\sqrt{2\Lambda} + \epsilon \cos \left(\sqrt{\frac{ 4\epsilon
(2\Lambda)^{3/2}}{3}} R\right),
\end{equation}
and take
\begin{equation}
\psi = \frac{4(2\Lambda)^{3/2}}{3} \epsilon^2 \tau
\end{equation}
in (\ref{RNS}), with $t\to i\tau$.
Then
\begin{equation}\label{vuc1}
V_d(r) = \frac{2(2\Lambda)^{3/2}}{3} \epsilon^3 \sin^2 (\sqrt{\frac{
4\epsilon(2\Lambda)^{3/2}}{3}} R) [\cos (\sqrt{\frac{
4\epsilon(2\Lambda)^{3/2}}{3
}} R) +1],
\end{equation}
and the metric in the limit $\epsilon \to 0$ is
\begin{equation} \label{uc1}
ds^2 = R^2 d\psi^2 + dR^2 + \frac{1}{2\Lambda} (d \theta^2 + \sin^2 \theta
d\phi^2).
\end{equation}
Thus this instanton has topology $S^2 \times R^2$. In fact, the metric
is the direct product of a flat $R^2$ and a round two-sphere of radius
$1/\sqrt{2\Lambda}$. The internal infinity $B^{\infty}$ is now at
$R=\infty$. We will take $R=R_0$ in the calculations, and then take
the limit $R_0 \to \infty$. The angle $\psi$ is interpreted as the
imaginary time. We could construct the same instanton from (\ref{chN})
in the limit $A \to 0$, by taking $\chi = \sqrt{A} R$. The gauge field
is
\begin{equation}
F =  \frac{1}{2\sqrt{\Lambda}} \sin \theta d\theta \wedge d\phi
\end{equation}
in the magnetic case, and becomes
\begin{equation}
F = - i \sqrt{\Lambda} R dR \wedge d\psi
\end{equation}
in the electric case. The Lorentzian metric obtained by taking $\psi
\to i \Psi$ represents Mink$^{(1,1)} \times S^2$ in Rindler
coordinates; the horizon at $R=0$ is the Rindler horizon.

Alternatively, we could define $x$ by
\begin{equation}
r = \frac{1}{\sqrt{2\Lambda}} +
\sqrt{\frac{2 (2\Lambda)^{3/2}}{3}}
\epsilon^{3/2} x,
\end{equation}
and take
\begin{equation}
\gamma = \sqrt{\frac{2 (2\Lambda)^{3/2}}{3}} \epsilon^{3/2} \tau.
\end{equation}
Then
\begin{equation}\label{vuc2}
V_d(r) = \frac{2\epsilon^3 (2\Lambda)^{3/2}}{3}\left( 1
+ \sqrt{\frac{2 (2\Lambda)^{3/2}}{3}}\epsilon^{1/2} x \right)^2
\left(1- \sqrt{\frac{2 (2\Lambda)^{3/2}}{3}} \epsilon^{1/2} x \right),
\end{equation}
and the metric in the limit $\epsilon \to 0$ is
\begin{equation} \label{uc2}
ds^2 = d\gamma^2 + dx^2 +  \frac{1}{2\Lambda} (d \theta^2 + \sin^2 \theta
d\phi^2).
\end{equation}
This instanton also has topology $S^2 \times R^2$, but the internal
infinity $B^{\infty}$ now has two components, $x=\pm \infty$. We will
evaluate the action for a region bounded by $x = \pm x_0$, and then
take $x_0 \to \infty$. It is $\gamma$ that is interpreted as the
analogue of imaginary time, and $\gamma$ runs from $-\infty < \gamma <
\infty$, so this looks just like flat space. The gauge field in this
case is
\begin{equation}
F =  \frac{1}{2\sqrt{\Lambda}} \sin \theta d\theta \wedge d\phi
\end{equation}
in the magnetic case, and
\begin{equation}
F = i \sqrt{\Lambda} d\gamma \wedge dx
\end{equation}
in the electric case. This solution describes the neighbourhood of a
point, when both horizons have receded to infinity. That is, the
Lorentzian section is just Mink$^{(1,1)} \times S^2$ in the usual
coordinates.

In summary: for sufficiently small mass, there are two solutions, the
lukewarm and cold solutions, which correspond to pair creation of
non-extreme and extreme black holes respectively. At given mass, the
cold solution has higher charge than the lukewarm solution. Once the
mass reaches $M = 1/(3\sqrt{\Lambda})$, there is a third solution, the
charged Nariai solution, which has lower charge than the other
two. When the mass reaches $M = 3/(4\sqrt{3\Lambda})$, the lukewarm
and charged Nariai solutions coincide, and there is no lukewarm
solution with higher mass. The cold and charged Nariai solutions
coincide, in the ultracold solution, when the mass reaches $M =
2/(3\sqrt{2\Lambda})$, and there are no regular solutions where the
mass is larger than this. The ratio of charge to mass is also at its
largest at this point, where $Q/M = 3/(2\sqrt{2})$. There is an
interesting analogy between the situation here and the Ernst
instantons; with the Ernst solution, there are also three ways in
which a non-singular instanton can be achieved. These are the
non-extreme (or type II) instanton, the extreme instanton, and the
type I instanton \cite{dgkt,dggh}. The lukewarm solution may be
thought of as the analogue of the non-extreme instanton, the cold
solution as the analogue of the extreme instanton, and the charged
Nariai solution as the analogue of the Type I instanton. The
plots of mass versus charge for the Reissner-Nordstr\"om de Sitter and
Ernst instantons are given in Figure 1 and Figure 2 respectively. While the
numerical values are not the same, the qualitative features of these
two plots are strikingly similar.

\begin{figure}
\vspace{65mm}
\includegraphics{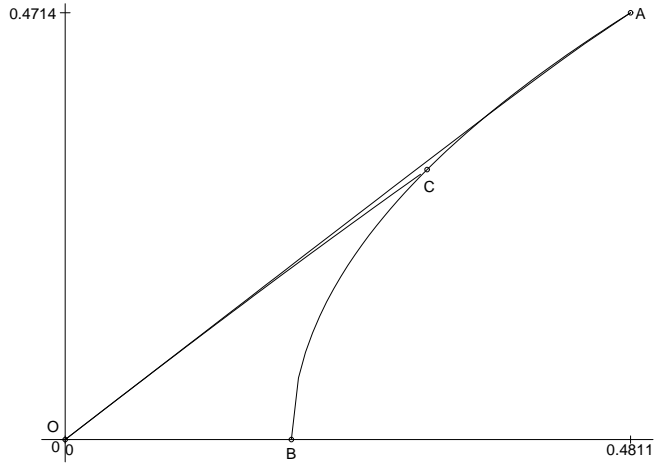}
\caption{The values of $Q$ and $M$ for which instantons can be
obtained in the electromagnetic case. The plot is of the dimensionless
quantities $q A$ vs. $m A$. The curve OA represents the extreme
solutions, OC represents the lukewarm solutions, and BA represents the
Type I instantons. The point at O is the Melvin solution. }
\end{figure}

\section{Pair creation rate and entropy}
\label{action}

In the previous section, we presented the Euclidean solutions which
provide the instantons for pair creation of charged black holes in a
cosmological background. We will now review their interpretation as
instantons, and use them to obtain approximate rates for these
processes.

The pair creation of non-extreme charged black holes in a cosmological
background is described by propagation from nothing to a surface
$\Sigma$ with topology $S^2 \times S^1$, since such a surface may be
thought of as a Wheeler wormhole attached to an $S^3$, and $S^3$ is
the topology of the spatial sections of de Sitter space. The pair
creation of extreme charged black holes is similarly described by
propagation to a surface with topology $S^2 \times R^1$. In this case
there is also a boundary component $B^\infty$, which represents an
``internal infinity''. We will use $\Sigma$ to denote the whole
boundary in this case, and $\Sigma_s$ to denote the part with topology
$S^2 \times R^1$. The amplitude for these processes will, at least
formally, be given by a path integral;
\begin{equation} \label{pint}
\Psi = \int d[g] d[A] e^{-I}.
\end{equation}
The integral is over all metrics and gauge fields which agree with the
given boundary data on $\Sigma$, and $\Psi$ may thus be thought of as
a functional of the boundary data.  We assume that, if there is a
Euclidean classical solution which interpolates within the given
boundary, then the  integral is dominated by the contribution from
it. That is, if there is an appropriate instanton, $\Psi$ will be
approximately
\begin{equation}
\Psi \approx e^{-I}, \label{V0}
\end{equation}
where $I$ is the action of this instanton. The partition function, and
thus the pair creation rate, is given by the square of this amplitude.

A spatial section of the lukewarm solution has topology $S^2 \times
S^1$, so half of the Euclidean section provides an instanton for the
pair creation of non-extreme charged black holes. We have to use half
of the solution, as we want the extrinsic curvature of $\Sigma$ to
vanish, so that $\Sigma$ can be interpreted as the zero-momentum
initial data for the Lorentzian extension. Similarly, half of the cold
solution can be used as an instanton to describe the pair creation of
extreme charged black holes. Note that these instantons only exist
when the data on $\Sigma$ specified in the path integral agrees with
the data induced by the solutions.

The whole of the Euclidean section in each case provides a ``bounce''
solution. In asymptotically-flat situations, one can deal with the
bounces rather than the instantons, but in a cosmological situation
this is no longer possible, as the boundary data on the surface
$\Sigma$ provide crucial information \cite{hawross}. We will therefore
be interested in the calculation of the action for the instantons.

We also need to consider what action we should use in the calculation
of (\ref{V0}). We want to use the action for which it is natural to
fix the boundary data on $\Sigma$ specified in the path integral
(\ref{pint}). That is, we want to use an action whose variation gives
the Euclidean equations of motion when the variation fixes these
boundary data on $\Sigma$ \cite{brown}. If we consider the action
(\ref{emac}), we can see that its variation will be
\begin{eqnarray}
\delta I &=& \mbox{ (terms giving the equations of motion) } \nonumber
\\ &&+ \mbox{
(gravitational boundary terms) } \nonumber \\
&& + \frac{1}{4\pi} \int_\Sigma d^3 x
\sqrt{h} F^{\mu\nu} n_\mu \delta A_\nu,
\end{eqnarray}
where $n_\mu$ is the normal to $\Sigma$ and $h_{ij}$ is the induced metric
on $\Sigma$ (see \cite{brown} for a more detailed discussion of the
gravitational boundary terms). Thus, the variation of (\ref{emac})
will only give the equations of motion if the variation is at fixed
gauge potential on the boundary, $A_i$. Note that it is not necessary
to fix the component $A_\mu n^\mu$ normal to the boundary.

For the magnetic Reissner-Nordstr\"om solutions, fixing the gauge
potential fixes the charge on each of the black holes, as the magnetic
charge is just given by the integral of $F_{ij}$ over a two-sphere
lying in the boundary. However, in the electric case, fixing the gauge
potential $A_i$ can be regarded as fixing a chemical potential
$\omega$ which is conjugate to the charge \cite{hawross}.  Holding the
charge fixed in the electric case is equivalent to fixing $n_\mu
F^{\mu i}$ on the boundary, as the electric charge is given by the
integral of the dual of $F$ over a two-sphere lying in
the boundary. Therefore, the appropriate action is
\begin{equation} \label{elac}
I_{el}  = I  - \frac{1}{4\pi} \int_{\Sigma} d^3 x \sqrt{h} F^{\mu\nu}
n_\mu A_\nu,
\end{equation}
as its variation is
\begin{eqnarray}
\delta I_{el} &=& \mbox{ (terms giving the equations of motion) }
\nonumber \\ &&+\mbox{
(gravitational boundary terms) } \nonumber \\
&&- \frac{1}{4\pi} \int_\Sigma d^3 x
\delta(\sqrt{h} F^{\mu\nu} n_\mu) A_\nu,
\end{eqnarray}
and so it gives the equations of motion when $\sqrt{h} n_\mu
F^{\mu i}$, and thus the electric charge, is held fixed.

Let us consider the pair creation where the electric and magnetic
charge are held fixed on $\Sigma$. Then the appropriate action will be
(\ref{emac}) in the magnetic case, and (\ref{elac}) in the electric
case. It is also worth pointing out that, since we identify $\Sigma$
($\Sigma_s$ in the cold case) with a surface of zero extrinsic curvature
in the Euclidean section, the gravitational boundary term in the
action (\ref{emac}) will make no contribution to the action. Thus the
action (\ref{emac}) for the instanton will just be half that of the
whole Euclidean section.

We will now consider each of the instantons derived in section
\ref{instsec}, and calculate the actions (\ref{emac}) in the magnetic
case and  (\ref{elac}) in the electric case. For the lukewarm solution we
have, in the magnetic case, $F^2 = 2Q^2/r^4$, and thus the action
(\ref{emac}) is \cite{moss}
\begin{eqnarray}
I^L &=&- \frac{\Lambda V^{(4)}}{8\pi} + \frac{1}{16\pi} \int d^4 x
\sqrt{ -g} F^2 \nonumber \\ &=& - \beta\Lambda \frac{r_4^3 - r_3^3}{12}+
\frac{Q^2}{4}\beta \left(\frac{1}{r_3} - \frac{1}{r_4}\right) =
-\frac{3\pi}{2 \Lambda} + \pi M \sqrt{\frac{3}{\Lambda}},
\end{eqnarray}
where $V^{(4)}$ is the four-volume of the instanton, and $\beta$ is
the period of $\tau$. In the electric
case, $F^2 = -2Q^2/r^4$, and we find that (\ref{emac}) gives
\begin{equation}
I^L_E = - \beta\Lambda \frac{r_4^3 - r_3^3}{12} - \frac{Q^2}{4}\beta
\left(\frac{1}{r_3} - \frac{1}{r_4}\right) = - \frac{3 \pi}{2 \Lambda}.
\end{equation}
To calculate the additional boundary term in (\ref{elac}), we have to
pick a gauge for the Maxwell field. To obtain a unique result, we have
to constrain the gauge choice to be regular at both horizons. A
suitable gauge choice for the lukewarm solution is
\begin{equation} \label{lgauge}
A = -i\frac{Q}{r^2}\tau dr.
\end{equation}
It might seem that this gauge choice involves a discontinuity at the
horizons, but in fact it does not. To consider whether there is a
discontinuity at the horizon, we should look at the gauge potential in
orthonormal coordinates. An orthonormal frame for the metric
(\ref{RNS}) is
\begin{equation}
{\bf e}_0 = V(r)^{1/2} dt, {\bf e}_1 = V(r)^{-1/2} dr, {\bf e}_2 = r
d\theta, {\bf e}_3 = r \sin \theta d\phi,
\end{equation}
and the gauge potential (\ref{lgauge}) is
\begin{equation}
A = - i V(r)^{1/2}\frac{Q}{r^2} \tau {\bf e}_1,
\end{equation}
which vanishes at $r=r_3$ and $r=r_4$.

To evaluate the  additional boundary term in (\ref{elac}), we take a
coordinate system such that the boundary is the surface
$\tau=0,\beta/2$ in the Euclidean section, and we take the integral in
the $r$ direction on the boundary to run from the black hole horizon
to the acceleration horizon along $\tau=0$, and back along
$\tau=\beta/2$. The additional term is
\begin{equation}
 \frac{1}{4\pi} \int_{\Sigma} d^3 x \sqrt{h} F^{\mu\nu} n_\mu A_\nu =
-\frac{Q^2}{2}\beta \left(\frac{1}{r_3} - \frac{1}{r_4}\right) = - \pi
M \sqrt{\frac{3}{\Lambda}}
\end{equation}
and thus the action (\ref{elac}) in the electric case is
\begin{equation}
I^L_{el} = I^L_E - \frac{1}{4\pi} \int_{\Sigma} d^3 x \sqrt{h}
F^{\mu\nu} n_\mu A_\nu = - \frac{3\pi}{2 \Lambda} + \pi M
\sqrt{\frac{3}{\Lambda}}.
\end{equation}
The relevant action is thus the same for the electric and magnetic
lukewarm instantons. It lies in the range $-3\pi/2\Lambda \leq I^L \leq
-3\pi/4\Lambda$, as $M < \sqrt{3/\Lambda}/4$.

For the cold solution, $F^2=2Q^2/r^4$ in the magnetic solution, and the
action (\ref{emac}) is
\begin{equation}
I^C = \beta\Lambda \frac{b^3 - \rho^3}{12} +  \frac{Q^2}{4}\beta
\left(\frac{1}{\rho} - \frac{1}{b}\right) = -\frac{\pi}{2} b^2.
\end{equation}
There is also an extrinsic curvature boundary term at $B^\infty$,
but this vanishes. In the electric case, $F^2 =
-2Q^2/r^4$, so the action (\ref{emac}) is
\begin{equation}
I^C_E = \beta\Lambda \frac{b^3 - \rho^3}{12} - \frac{Q^2}{4}\beta
\left(\frac{1}{\rho} - \frac{1}{b}\right) = -\frac{\pi}{2} b^2 -
\frac{Q^2}{2}\beta \left(\frac{1}{\rho} - \frac{1}{b}\right).
\end{equation}
A suitable gauge potential, which is regular everywhere on the instanton, is
\begin{equation}
A = -iQ\left(\frac{1}{r} - \frac{1}{b}\right)d\tau.
\end{equation}
The integral over $\Sigma$ now consists of two parts; there is an
integral over the $S^2 \times R^1$ factor, which we take to be from
$r=\rho + \epsilon$ to the acceleration horizon at $r=b$ along $\tau =
0$, and back along $\tau = \beta/2$, and an integral over the internal
infinity $r=\rho+\epsilon$, which is in the direction of {\em
decreasing} $\tau$. The additional boundary term in (\ref{elac}) is
\begin{equation}
\frac{1}{4\pi} \int_{\Sigma} d^3 x \sqrt{h} F^{\mu\nu} n_\mu A_\nu =  -
\frac{Q^2}{2}\beta \left(\frac{1}{\rho} - \frac{1}{b}\right),
\end{equation}
and thus (\ref{elac}) is
\begin{equation}
I^C_{el} = -\frac{\pi}{2} b^2.
\end{equation}
Again, the action is the same in the electric and magnetic cases.
It lies in the range $-3\pi/2\Lambda \leq I^C \leq  - \pi/4\Lambda$.

The actions in the charged Nariai case have already been computed in
\cite{hawross}. In the magnetic case, $F^2 = 2Q^2/B^2$, and the action
(\ref{emac}) is
\begin{equation}
I^{CN} = -\frac{\pi}{B}.
\end{equation}
In the electric case, $F^2 = -2Q^2/B^2$, so the action (\ref{emac}) is
\begin{equation}
I^{CN}_E = -\frac{\pi}{A}.
\end{equation}
A suitable gauge potential is
\begin{equation}
A = i\frac{QB}{A}\sin(\chi)\psi d\chi.
\end{equation}
We take the boundary to be the surface $\psi=0, \psi=\pi$, and
integrate from the black hole horizon ($\chi=\pi$) to the acceleration
horizon ($\chi=0$) along $\psi=0$, and back along $\psi=\pi$, so the
additional boundary term in (\ref{elac}) is
\begin{equation}
\frac{1}{4\pi} \int_{\Sigma} d^3 x \sqrt{h} F^{\mu\nu} n_\mu A_\nu =
-\frac{1}{4\pi} \frac{Q^2 B}{A} \int \psi \sin \chi \sin \theta d\chi
d\theta d\phi = -2\pi Q^2 \frac{B}{A},
\end{equation}
and thus (\ref{elac}) is
\begin{equation}
I^{CN}_{el} = -\frac{\pi}{A} + \frac{2\pi Q^2 B}{A} = -\frac{\pi}{B}.
\end{equation}
The relevant action is the same in the electric and magnetic cases,
and it lies in the range $-\pi/\Lambda \leq I^{CN} \leq
-\pi/2\Lambda$.

For both metrics which can be constructed in the ultracold case, $F^2
= 2\Lambda$ in the magnetic solution, so the volume contribution to
the action (\ref{emac}) vanishes. Let us consider first the metric
(\ref{uc1}). Then
\begin{equation}
I^{UC1} = -\frac{1}{8\pi} \int_{B^\infty} \sqrt{h} K = -\frac{\pi}{4\Lambda}.
\end{equation}
In the electric solution, $F^2 = -2\Lambda$,
so (\ref{emac}) gives
\begin{equation}
I^{UC1}_E = - \frac{\Lambda V^{(4)}}{4 \pi}-\frac{1}{8\pi}
\int_{B^\infty} \sqrt{h} K = - \pi R_0^2/4  -\frac{\pi}{4\Lambda}
\end{equation}
(the boundary $B^{\infty}$ in this case is the surface $R=R_0$). One
could take the electric gauge potential to be
\begin{equation}
A = - \frac{i}{2}\sqrt{\Lambda} R^2 d\psi.
\end{equation}
We define $\Sigma$ to be the surfaces $\psi = 0, \psi = \pi$,
together with the semi-circle at $R=R_0$ lying between them, and take the
integral around the boundary to be from $R=R_0$ to $R=0$ along $\psi=0$, back
along $\psi=\pi$, and around $R=R_0$ in the direction of decreasing
$\psi$. The additional boundary term in (\ref{elac}) is
\begin{equation}
\frac{1}{4\pi} \int_{\Sigma} d^3 x \sqrt{h} F^{\mu\nu} n_\mu A_\nu = -\pi
R_0^2/4,
\end{equation}
so (\ref{elac}) is
\begin{equation}
I^{UC1}_{el} =  -\frac{\pi}{4\Lambda}.
\end{equation}
This action agrees with the limit of the action of the cold solution
as it approaches ultracold.

If we consider instead the metric (\ref{uc2}), the action vanishes in
the magnetic case, $I^{UC2}=0$, as the extrinsic curvature surface term
at $x=\pm x_0$ vanishes as well. In the electric case, $F^2 =
-2\Lambda$, so if we consider the action for the region between two
surfaces $\gamma=\pm \gamma_0$, (\ref{emac}) gives
\begin{equation}
I^{UC2}_E =  - \frac{\Lambda V^{(4)}}{4 \pi} = - 2x_0 \gamma_0.
\end{equation}
One could take the electric gauge potential to be
\begin{equation}
A = -i \sqrt{\Lambda} x d\gamma.
\end{equation}
The additional boundary term in (\ref{elac}) is then
\begin{equation}
\frac{1}{4\pi} \int_{\Sigma} d^3 x \sqrt{h} F^{\mu\nu} n_\mu A_\nu = -2 x_0
\gamma_0,
\end{equation}
so (\ref{elac}) vanishes as well, $I^{UC2}_{el} =0$.

One can use the action of an instanton to approximate the wavefunction
for the propagation to some final surface, so the action gives the
approximate amplitude at which this process occurs. One can square
this to get the rate.  The actions we have just calculated for the
cold and lukewarm instantons give the rate for pair creation of black
holes in a cosmological background by these instantons. The pair
creation rate is approximately
\begin{equation}
\Gamma = \Psi^2 \approx e^{-2I},
\end{equation}
in each case, where $I$ is the relevant action. Since de Sitter space
has action $I_{\rm de\ Sitter} = -3\pi/2\Lambda$, which is the lower
bound of the action for the cold and lukewarm instantons, the rate at
which black hole pairs are created relative to the rate at which de
Sitter space is itself created is less than one. That is, the pair
creation of black holes is suppressed.  The situation is illustrated
in Figure 3.

\begin{figure}
\vspace{65mm}
\includegraphics{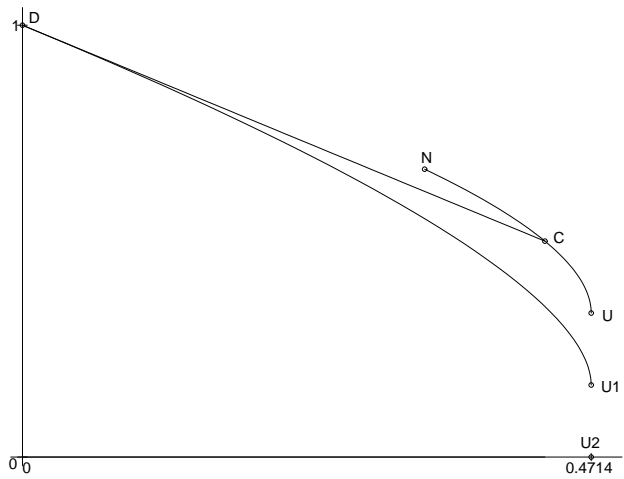}
\caption{The action for the various instantons in the cosmological
case. The action as a fraction of the action for de Sitter space,
$I/I_{\rm de\ Sitter}$, is plotted against the dimensionless mass
$M\protect\sqrt{\Lambda}$. The curve DU1 represents the cold solutions,
DC represents the lukewarm solutions, and NU represents the charged
Nariai solutions. The point at D is de Sitter space,  N is the Nariai
solution, and U1 and U2 represent the actions of the first and second
type of ultracold solutions. Note that U does {\em not} correspond to
one of the ultracold solutions.}
\end{figure}

Since the instantons are all cosmological solutions, there is no
asymptotic region in the Euclidean section (that is, there is no
`point at infinity'). This can be interpreted as meaning that these
solutions are closed systems, and thus necessarily have fixed
energy. They should therefore be interpreted as a contribution to the
microcanonical ensemble, as this is the thermodynamical ensemble at
fixed energy.  The partition function $Z = \Psi^2$ should therefore be
interpreted as the density of states, and thus the entropy will be
just the ln of this partition function, $S = \ln Z$
\cite{brown,hawross}.  The contribution to the entropy from the
instantons is thus just
\begin{equation}
S = -2 I.
\end{equation}
As we might expect, the entropy turns out to be just a quarter of the
total area of the horizons which appear in the instanton (which we
denote by ${\cal A}$). In the lukewarm case, there are two horizons,
at $r=r_3$ and $r=r_4$, so ${\cal A}/4 = \pi r_3^2 + \pi r_4^2$, but
\begin{equation}
r_{3,4} = \frac{1}{2} \left[ \sqrt{\frac{3}{\Lambda}} \pm
\sqrt{\frac{3}{\Lambda} - 4M \sqrt{\frac{3}{\Lambda}}} \right],
\end{equation}
so
\begin{equation}
{\cal A}/4 = \frac{3\pi}{\Lambda} - 2\pi M \sqrt{\frac{3}{\Lambda}},
\end{equation}
and thus
\begin{equation}
S^L = -2 I^{L}= {\cal A}/4.
\end{equation}
In the cold case, only the acceleration horizon at $r=b$ is part of
the instanton, and it has area ${\cal A} = 4\pi b^2$, so
\begin{equation}
S^C = -2 I^C = \pi b^2 = {\cal A}/4.
\end{equation}
In the charged Nariai case, there
are again two horizons, which both have area $4\pi/B$. Thus ${\cal
A}/4 = 2\pi/B$, and
\begin{equation}
S^{CN} = -2 I^{CN} = 2\pi/B ={\cal A}/4.
\end{equation}
For the first type of ultracold solution, (\ref{uc1}), the
surface $R=0$ is interpreted as a Rindler horizon, which has area
${\cal A} = 2\pi/\Lambda$. Thus,
\begin{equation}
S^{UC1} = -2 I^{UC1} = \pi/2\Lambda = {\cal A}/4.
\end{equation}
For the other type of ultracold solution, there are no
horizons, and the entropy vanishes, $S^{UC2} = -2 I^{UC2} =0$, as we
expect. Thus we see that horizons contribute to the gravitational
entropy only if they are in the instanton; in particular, extreme
black hole horizons make no contribution to the entropy, even if they
have non-zero area, as discovered in \cite{entar}.

Thus, the usual relation between the entropy and the area of
the horizons extends to all these solutions. Just as for the Ernst
instantons, the pair creation of extreme black holes is suppressed
relative to the pair creation of non-extreme black holes by a factor of
$e^{S_{bh}}$, where $S_{bh}$ is the entropy associated with the black
hole horizon.

\section{Conclusions}
\label{concl}

The pair creation of charged black holes by a strong electromagnetic
field has been a subject of considerable recent interest.  We have
seen that charged black holes can also be pair created in a background
with a positive cosmological constant. The nature of the instantons
describing this pair creation is very similar to that of the
instantons describing pair creation in an electromagnetic field. There
is a non-extreme and an extreme instanton. As in the electromagnetic
case, the pair creation of extreme black holes is suppressed relative
to that of non-extreme black holes by a factor of $e^{S_{bh}}$, where
$S_{bh}$ is the entropy associated with the black hole horizon. This
is further evidence that $e^{S_{bh}}$ should be regarded as the number
of internal states of the black holes.

The pair creation rate obtained from the instantons is in fact just
$e^S$, where $S$ is the total gravitational entropy of the instanton,
that is, a quarter the area of the black hole and the cosmological
horizons. Seen from this point of view, black hole pair creation in de Sitter
space is suppressed simply because de Sitter space has a higher
entropy; that is, the single horizon of de Sitter space has an area
larger than the combined area of the horizons in the instantons. This
is also similar to the electromagnetic case, where the pair creation
rate was $e^{\Delta {\cal A}/4 + {\cal A}_{bh}/4}$ in the non-extreme
case, with the suppression being due to the fact that the difference
in acceleration horizon area $\Delta {\cal A}$ was negative.

Because the conical singularities in the charged $C$ metric with a
cosmological constant can only be eliminated when it reduces to the
Reissner-Nordstr\"om de Sitter metric, we are fairly confident that
the instantons we have described in this paper are the only ones which
can be interpreted as representing pair creation of charged black
holes in a cosmological background. It might be interesting to see if
these solutions could be extended to dilaton gravity with some kind of
effective cosmological constant, as the presence of the dilaton field
might allow some more possibilities.

We have also observed that the Type I instanton discovered in
\cite{dgkt}, where the conical singularities in the ordinary charged
$C$ metric are eliminated by allowing $\xi_2$ and $\xi_3$ to coincide,
is in fact the Reissner-Nordstr\"om metric.

\section{Acknowledgements}

R.B.M. is grateful for the hospitality of D.A.M.T.P., where this work
was performed, and for the support of the Natural Sciences and
Engineering Research Council of Canada.  S.F.R. thanks Raphael Bousso
for illuminating discussions, and the Association of Commonwealth
Universities and the Natural Sciences and Engineering Research Council
of Canada for financial support.

\end{document}